\documentclass[aps,prl,twocolumn,superscriptaddress]{revtex4-1}%
\usepackage{epsf,epsfig}
\usepackage[utf8]{inputenc}
\usepackage{amssymb,amsmath,amsfonts}
\usepackage{color}
\usepackage{slashed}
\usepackage{latexsym}
\usepackage{siunitx}
\usepackage{braket}
\usepackage[colorlinks=true]{hyperref}  
\hypersetup{
    bookmarks=true,         
    unicode=false,          
    pdftoolbar=true,        
    pdfmenubar=true,        
    pdffitwindow=false,     
    pdfstartview={FitH},    
    colorlinks=true,       
    linkcolor=magenta, 
    citecolor=blue,        
    filecolor=magenta,      
    urlcolor=cyan           
} 

\newcommand{\bea}{\begin{eqnarray}}
\newcommand{\eea}{\end{eqnarray}}
\newcommand{\ba}{\begin{eqnarray}}
\newcommand{\ea}{\end{eqnarray}}

\newcommand{\beq}{\begin{equation}}
\newcommand{\eeq}{\end{equation}}
\newcommand{\beqa}{\begin{eqnarray}}
\newcommand{\eeqa}{\end{eqnarray}}
\newcommand{\beqar}{\begin{eqnarray*}}
	\newcommand{\eeqar}{\end{eqnarray*}}

\newcommand{\be}{\begin{equation}}
\newcommand{\ee}{\end{equation}}

\newcommand{\eg}{{\it e.g.,}\ }

\usepackage{color}

\begin{document}

\title{Holographic Complexity and Thermodynamic Volume}

%

\author{Abdulrahim Al Balushi}
\email{a2albalu@uwaterloo.ca}
\affiliation{Department of Physics and Astronomy, University of Waterloo, Waterloo, Ontario, N2L 3G1, Canada}

\author{Robie A. Hennigar} 
\email{rhennigar@mun.ca}
\affiliation{Department of Mathematics and Statistics, Memorial University of Newfoundland, St. John's, Newfoundland and Labrador, A1C 5S7, Canada}

\author{Hari K. Kunduri}
\email{hkkunduri@mun.ca}
\affiliation{Department of Mathematics and Statistics, Memorial University of Newfoundland, St. John's, Newfoundland and Labrador, A1C 5S7, Canada}

\author{Robert B. Mann}
\email{rbmann@uwaterloo.ca}
\affiliation{Department of Physics and Astronomy, University of Waterloo,
	Waterloo, Ontario, N2L 3G1, Canada}


\begin{abstract}

We study the holographic complexity conjectures for rotating black holes, uncovering a relationship between the complexity of formation and the thermodynamic volume of the black hole. We suggest that it is the thermodynamic volume and not the entropy that controls the complexity of formation of large black holes in both the Complexity Equals Action and Complexity Equals Volume proposals in general. Our proposal reduces to known results involving the entropy in settings where the thermodynamic volume and entropy are not independent, but has broader scope. Assuming a conjectured inequality is obeyed by the thermodynamic volume, we establish that the complexity of formation is bounded from below by the entropy for large black holes.

\end{abstract}
\maketitle

In recent years there has been dramatic progress in understanding the connections between gravity and quantum information. The quintessential example of this is entanglement in the context of the AdS/CFT correspondence. Through the Ryu-Takayanagi prescription and its generalizations~\cite{Ryu2006,Hubeny:2007xt, Casini2011, Lewkowycz:2013nqa} the duality relates entanglement between spacetime regions in the field theory to the existence of minimal surfaces in the bulk, a situation often described by the slogan ``entanglement$=$geometry''. 


Recently it has been suggested that entanglement may not be sufficient to fully describe physics in extreme regimes, such as the late-time dynamics of black holes~\cite{Susskind:2014rva, Susskind:2014moa}, and that instead
 \textit{complexity} of a dual CFT state provides information that entanglement does not. Roughly speaking, complexity provides a measure of how difficult it is to construct certain states in the theory starting from simple unentangled states using a fixed set of universal gates. While well-established in quantum mechanics, circuit complexity in quantum field theory is an area of active investigation, and there remains much to understand about its role in the holographic dictionary~\cite{Chapman2017, Jefferson2017}. There have been a number of proposals suggesting how the complexity of the field theory state should be expressed in terms of bulk observables. The two most well-studied of these proposals are the ``Complexity$=$Volume'' (CV)~\cite{Stanford2014}  and ``Complexity$=$Action'' (CA)~\cite{Brown2016a,Brown2016} conjectures. The former relates complexity to the volume of extremal codimension-one surfaces in the bulk, while the latter relates complexity to the value of the gravitational action on a region of spacetime known as the Wheeler-DeWitt (WDW) patch.\footnote{See, {\it e.g.},~\cite{Bernamonti:2019zyy, Bernamonti:2020bcf} for preliminary investigations into the connections between the circuit complexity and holographic complexity proposals.}   

A number of properties of complexity as defined by the CV and CA proposals are now well-understood for black holes, with both proposals generally 
yielding qualitatively similar results, but not always~\cite{Carmi:2016wjl, Chapman:2018lsv, Chapman:2018bqj, Andrews:2019hvq, Bernamonti:2019zyy, Bernamonti:2020bcf}. For example, in both proposals it is known that at late times the complexity grows linearly in time at a rate characterized by the mass, or other thermodynamic potentials, of the black hole~\cite{Brown2016, Cai2016, Huang:2016fks, Carmi2017, Cano2018}. In both proposals the response of complexity to perturbations follows the ``switchback effect''~\cite{Chapman:2018dem, Chapman:2018lsv}.  Most relevant for us here is the finding of~\cite{Chapman2017} (see also~\cite{Carmi2017}) that in both proposals the complexity of formation of large, (un)charged static, and spherically symmetric black holes is proportional to the black hole entropy.

Here we report on the first investigation of complexity for rotating black holes. From a holographic perspective, rotating black holes are dual to thermofield double states living on a rotating spacetime~\cite{Hawking-rotation, Hawking:1999dp, Caldarelli:1999xj}. However, our main motivation here is to exploit the more complicated  geometric structure of rotating black holes to test the complexity proposals for universal and divergent features that may not be evident in simpler geometries.  The approach of understanding the behaviour of an observable under deformations of the state or theory (\eg through the addition of higher-curvature terms in the action) has been a fruitful line of investigation for identifying universal relationships and testing conjectures in the context of AdS/CFT~\cite{Brigante:2007nu, Myers:2010jv, Myers:2010tj, Mezei:2014zla, Bueno:2015rda, Bueno:2018yzo}.

We shall exploit the observation that the causal structure of a class of odd-dimensional rotating black holes is far simpler than the general situation. This allows for computations that would be effectively intractable in the general situation to be carried out largely analytically.  Remarkably we find a connection  between the complexity of formation and the \textit{thermodynamic volume} $V$ of the black hole, indicating that it is this quantity and not the entropy that governs its behaviour in  both the CV and CA proposals.  In the static limit, we recover previously known results.

Thermodynamic volume is a quantity that arises naturally when generalizing the Komar definition of mass from asymptotically flat spacetimes to those with (A)dS asymptotics and plays a central role in extending Smarr's formula from flat spacetimes to AdS spacetimes~\cite{Kastor:2009wy, Cvetic:2010jb}. This extended Smarr relation reads
\begin{equation} \label{smarr}
(D-3)M = (D-2)TS + (D-2) \Omega_i J^i + \frac{\Lambda V}{4 \pi G_N}  \, ,
\end{equation}
where $D$ is the spacetime dimension, $T$ is the Hawking temperature, $S$ is the entropy, $\Omega_i$ are the horizon angular velocities, $J^i$ are the independent angular momenta, $\Lambda\equiv -(D-1)(D-2)/2\ell^2$ is the cosmological constant, and $\ell$ is the AdS length scale. If one allows for variations in the cosmological constant, the thermodynamic volume appears as the conjugate quantity to variations in $\Lambda$:
\begin{equation}\label{flaw}
\delta M = T \delta S + \Omega_i \delta J^i - \frac{V \delta \Lambda}{8 \pi G_N}  \,.
\end{equation}
Interpreting $P \equiv -\Lambda/(8 \pi G_N)$ as a pressure, the form of the first law appearing above identifies the mass as the \textit{enthalpy} of spacetime, rather than the internal energy.  In general $S$ and $V$ are independent quantities~\cite{Cvetic:2010jb}, but in certain cases (for example Reissner-Nordstrom-AdS) 
they both depend on a single parameter, with $S \sim V^{(D-2)/(D-1)}$.
  The implications of the thermodynamic volume have been extensively explored in the gravitational context --- see~\cite{Kubiznak:2016qmn} for a recent review --- but its role in holography remains comparatively unexplored (though see~\cite{Johnson:2014yja, Kastor:2014dra, Karch:2015rpa, Caceres:2016xjz, Sinamuli:2017rhp, Johnson:2018amj, Johnson:2019wcq, Rosso:2020zkk} for progress on this front).

There have been already a number of attempts to connect thermodynamic volume to the idea of complexity.  There is a sense in which this is natural --- in many situations the thermodynamic volume is related to the spacetime volume inside the black hole~\cite{Cvetic:2010jb, Bordo:2020ryp}, which is precisely what complexity is designed to probe. However, these investigations have either invoked new proposals for complexity~\cite{Couch:2016exn, Fan:2018wnv}, or re-expressed known results in terms of the thermodynamic volume for interpretational reasons~\cite{Huang:2016fks, Liu:2019mxz,Sun:2019yps}. Our result is the first to show concretely that thermodynamic volume emerges naturally and unambiguously in both the original CV and CA proposals in a way wholly distinct from entropy.


  
\textbf{Solutions and global structure}.---The Myers-Perry-AdS black hole solutions in $D = 2N + 3$ odd dimensions are characterized by their mass and $N + 1$ independent angular momenta $J_i$~\cite{Gibbons:2004uw}. In the special case where all angular momenta are equal,  considerable simplification occurs. The metric depends only on the radial coordinate and the line element reads~\cite{Kunduri:2006qa}
\begin{align}
ds^2 =& -f(r)^2 dt^2 + g(r)^2 dr^2 + h(r)^2 \left[ d\psi +  A - \Omega(r) dt\right]^2 
\nonumber\\
 &+ r^2 \hat{g}_{ab} dx^a dx^b
\end{align}
where
\begin{align}
g(r) ^2 =& \left( 1+ \frac{r^2}{\ell^2} - \frac{2m \Xi}{r^{2N}} + \frac{2m a^2}{r^{2N + 2}}\right)^{-1} ,
\nonumber\\
h(r)^2 =& r^2 \left( 1 + \frac{2m a^2}{r^{2N+2}}\right), \qquad \Omega(r) = \frac{2ma}{r^{2N} h^2}, 
\end{align} and
\begin{equation}
f(r) = \frac{r}{g(r) h(r)}, \qquad \Xi = 1 - \frac{a^2}{\ell^2}. 
\end{equation}
The metric $\hat{g}$ is the Fubini-Study metric on $\mathbb{CP}^N$ with curvature normalized so that $\hat{R}_{ij} = 2 (N+1) \hat{g}_{ij}$ and $A$ is a 1-form on $\mathbb{CP}^N$ that satisfies $dA =2J$ where $J$ is the K\"ahler form.  The basic example is in $D=5$, in which case $N=1$ and we have $\mathbb{CP}^1 \cong S^2$ with the metric
\begin{equation}
\hat{g} = \frac{1}{4} \left( d\theta^2 + \sin^2 \theta d\phi^2 \right), \qquad A = \frac{1}{2} \cos \theta d\phi \, .
\end{equation}
The asymptotic region is obtained in the limit $r \to \infty$, where we recover the  usual AdS$_{2N+3}$ metric provided we periodically identify $\psi \sim \psi + 2\pi$. 

The spacetime contains a horizon at $r = r_+$ where $r_+$ is the largest root of $g^{-2}(r_+) = 0$. The hypersurface $r = r_+$ is a smooth Killing horizon with null generator 
\begin{equation}
\xi = \frac{\partial}{\partial t} + \Omega_H \frac{\partial}{\partial \psi} , \qquad \Omega_H = \frac{2 m a}{r_+^{2N + 2} + 2m a^2} \, .
\end{equation}  
There is also an inner Cauchy horizon at $r = r_-$ which is the smaller of the two positive real roots of $g^{-2}(r)$. 

The conserved charges corresponding to mass and angular momentum are \cite{Gibbons:2004uw,Gibbons:2004ai} 
\begin{equation}\label{thermoMass}
M = \frac{\Omega_{2N+1} m }{4\pi G_N} \left( N + \frac{1}{2} + \frac{a^2}{2\ell^2}\right) , \quad J = \frac{\Omega_{2N+1}}{4\pi G_N} (N+1) m a \, ,
\end{equation} where $\Omega_{2N+1} = 2 \pi^{N+1}/\Gamma(N+1)$  
is the area of a unit $2N+1$ sphere.  We emphasize that the single angular momentum $J$ corresponds to equal angular momenta $J_i = J/(N+1)$ in each of the $N+1$-orthogonal planes of rotation. The black hole's entropy and temperature are given by
\begin{align}
S &=\frac{\Omega_{2N+1}h(r_+)r_+^{2N}}{4G_N}, \label{entropy}
\\
T &=\frac{1}{2\pi h(r_+)} \left[(N+1) \left(1 + \frac{r_+^2}{\ell^2}\right) - \frac{\ell^2 r_+^2}{(r_+^2-a^2)\ell^2 - r_+^2 a^2} \right]\, , \nonumber
\end{align}
while the thermodynamic volume is \cite{Altamirano:2014tva}
\begin{equation}\label{vol}
V = \frac{r_+^{2(N+1)} \Omega_{2N+1}}{2(N+1)} + \frac{4 \pi a J}{(2N+1)(N+1)} \, .
\end{equation} 
Note in particular that the entropy and thermodynamic volume are independent functions of $r_+$ and $r_-$ (or $m$ and $a$).
Within the framework of extended thermodynamics, the thermodynamic volume is conjugate to the pressure
\begin{equation}\label{press}
P = -\frac{\Lambda}{8\pi G_N} = \frac{(N+1)(2N + 1)}{8\pi \ell^2G_N}  \, .
\end{equation} 
These thermodynamic quantities satisfy the extended Smarr relation~\eqref{smarr} and first law~\eqref{flaw}. 

 The entropy presents two different scaling regimes, depending on whether the black hole is close to extremality or close to the static limit. For large black holes these regimes are characterized by the scaling
\be\label{entScale} 
S \underset{\frac{r_-}{r_+} \to 0}{\sim} \left(\frac{r_+}{\ell} \right)^{2N+1} \quad \text{and} \quad S \underset{\frac{r_-}{r_+} \to 1}{\sim} \left(\frac{r_+}{\ell} \right)^{2N+2} \, .
\ee
This should be contrasted with the scaling of $V$ in the same regimes, which satisfies
\be\label{volScale} 
V \underset{\frac{r_-}{r_+} \to 0}{\sim} \left(\frac{r_+}{\ell} \right)^{2N+2} \quad \text{and} \quad V \underset{\frac{r_-}{r_+} \to 1}{\sim} \left(\frac{r_+}{\ell} \right)^{2N+4} \, .
\ee
In the static limit $r_-/r_+ \to 0$ the scaling of the entropy and volume is related by $S \sim V^{(D-2)/(D-1)}$,  the same relationship that holds generally for the Schwarzschild-AdS and Riessner-Nordstr{\"o}m-AdS black holes. 

\begin{figure}
\centering
\includegraphics[width=0.3\textwidth]{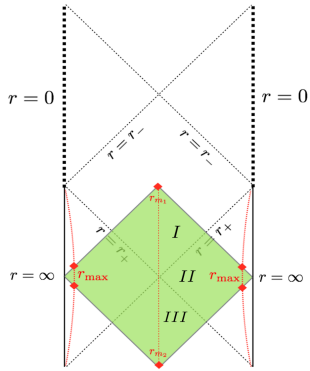}
\caption{A Penrose diagram for the equal-spinning Myers-Perry-AdS spacetime. The shaded green region represents the WDW patch. The full diagram is an infinite strip comprised of infinite repetition of the segment shown here.}
\label{penrose}
\end{figure}

Due to the enhanced symmetry of the equal-spinning solution, its causal structure is qualitatively similar to that of the Riessner-Nordstr{\"o}m-AdS solution, as can be confirmed by an analysis of the lightcone structure~\cite{Pretorius1998, AlBalushi2019}. Unlike the general situation for rotating black holes where $r = 0$ represents a ``ring singularity'' that can be traversed, the timelike surface $r = 0$ in these metrics is totally singular. We show in Fig.~\ref{penrose} a Penrose diagram for the spacetime, including also the WDW patch.

\textbf{Complexity equals volume}.---Let us consider now the complexity of formation within the CV proposal. According to the CV proposal, the complexity of a holographic state at the boundary time slice $\Upsilon$ is related to the volume of an extremal codimension-one slice $\mathcal{B}$ by
\begin{equation}
\mathcal{C}_\mathcal{V}(\Upsilon)=\max\limits_{\Upsilon=\partial\mathcal{B}}\left[\frac{\mathcal{V}(\mathcal{B})}{G_NR}\right] \, ,
\end{equation}
where $R$ is an arbitrary length scale. Since we are interested in the complexity of formation, we consider the $t = 0$ timeslice of the boundary and subtract from this the analogous result for the AdS vacuum.

We take as coordinates on a codimension-one surface $(\lambda, \vec{\Omega})$ where $\vec{\Omega}$ denotes the angular coordinates of the metric. Writing the metric in ingoing coordinates $(r, v)$ and parameterizing $r = r(\lambda)$ and $v = v(\lambda)$ \footnote{This choice is possible only because of the enhanced symmetry of the metric. For a general rotating black hole these functions would depend also on the polar angles.}, it is straight-forward to show that the volume functional is 
\begin{align}\label{MP-V}
\mathcal{V}&=2\Omega_{D-2}\int d\lambda\ h(r)r^{D-3}\sqrt{-f(r)^2\dot{v}^2+2g(r)f(r)\dot{v}{\dot{r}}} \, .
\end{align}
Stationary points of this functional represent surfaces of extremal volume, while the volume of those surfaces is then obtained by evaluating~\eqref{MP-V} on-shell. Straight-forward computations~\cite{UsFuture} allow us to deduce that
\begin{align}\label{extVol}
\mathcal{V}&=2\Omega_{D-2}\int_{r_+}^{r_{\rm max}}dr \, r^{(D-3)}h(r) g(r) \, ,
\end{align}
for the $t = 0$ timeslice.
Here the integration is cutoff at some large but finite value $r_{\rm max}$. The complexity of formation is obtained by subtracting from~\eqref{extVol} the analogous volume for two copies of the AdS vacuum \footnote{It is straightforward to show~\cite{UsFuture} that differences in the definition of $r_{\rm max}$ between the black hole spacetime and global AdS do not contribute to the integral.}:
\be 
\mathcal{V}_{\rm AdS} = \Omega_{D-2} \int_{0}^{r_{\rm max}} dr \frac{r^{D-2} }{\sqrt{1 + r^2/\ell^2}} \, ,
\ee 
and then taking the limit $r_{\rm max} \to \infty$.  This yields  
\be 
\Delta \mathcal{C}_\mathcal{V} = \lim_{r_{\rm max} \to \infty} \frac{\left[ \mathcal{V} - 2 \mathcal{V}_{AdS} \right]}{G_N R} \, . 
\ee

\begin{figure}
\centering
\includegraphics[width=0.4\textwidth]{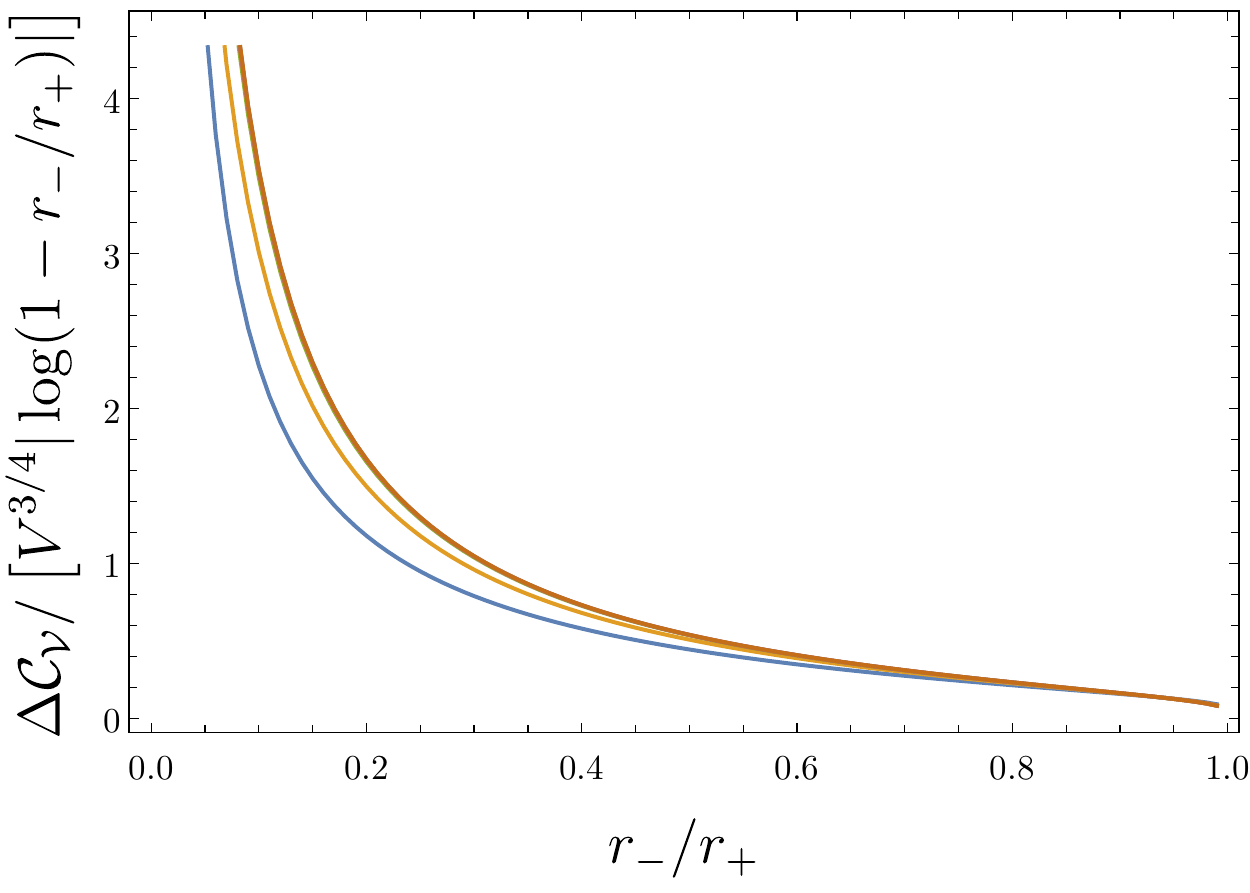}
\caption{A plot showing the CV complexity of formation normalized by the thermodynamic volume as a function of the ratio $r_-/r_+$ in five dimensions. The plot shows curves for fixed $r_+/\ell = 10, 10^2, 10^3, 10^4, 10^5, 10^6$ and $10^7$, however after $r_+/\ell = 1000$ the curves are visually indistinguishable.}
\label{CVplot}
\end{figure}

To understand the behaviour of $\Delta \mathcal{C}_\mathcal{V}$ for large black holes,   we plot it as a function of $r_-/r_+$ in  Fig.~\ref{CVplot} in $D=5$  for several different values of $r_+/\ell$. As extremality is approached, the complexity of formation exhibits a logarthmic divergence~\cite{UsFuture}, similar to what occurs for charged black holes~\cite{Carmi2017}. In the plot, we have normalized $ \Delta \mathcal{C}_\mathcal{V} $ taking into account this divergence, and have also normalized by the thermodynamic volume to an appropriate power. While the plot shows the curve for seven distinct values of $r_+/\ell$, only three curves are actually distinguishable. This illustrates that, for large black holes, the scaling of the complexity of formation with $r_+/\ell$ matches the scaling of the thermodynamic volume.

\begin{table}
\centering
\begin{ruledtabular}
\begin{tabular}{ccc} 
{Dimension} & {$\beta$ such that $\Delta \mathcal{C}_\mathcal{V} \sim (r_+/\ell)^\beta$} & {$V^{(D-2)/(D-1)}$} \\ 
5 & {$4.50000$} & {$9/2 = 4.5$}  \\
7 & {$6.66667$} &  {$20/3 \approx 6.66667$} \\
9 & {$8.75000$} & {$35/4 = 8.75$} \\
11 & {$10.80000$}& {$54/5 = 10.8$}  \\ 
13 & {$12.83333$} & {$77/6 \approx 12.83333$}  \\
15 & {$14.85714$} & {$104/7 \approx 14.85714 $} \\
17 & {$16.87500$} & {$135/8 \approx 16.87500 $} \\
19 & {$18.88889$} &{$170/9 \approx 18.88889$} \\ 
21 & {$20.90000$}& {$209/10 = 20.9$} \\
23 & {$22.90909$} & {$252/11 \approx 22.90909 $}   \\
25 & {$24.91667$} &  {$299/12 \approx 24.91667$}\\
27 & {$26.92308$} & {$350/13 \approx 26.92308$} \\
\end{tabular}
\end{ruledtabular}
\caption{Table comparing scaling of $\Delta \mathcal{C}_\mathcal{V}$ with the scaling of the thermodynamic volume $V^{(D-2)/(D-1)}$ for large $r_+/\ell$. The $\Delta \mathcal{C}_\mathcal{V}$ data is obtained numerically by evaluating the complexity of formation between $r_+/\ell = 10^{10}$ and $r_+/\ell = 10^{20}$ and we work close to extremality with $r_-/r_+ =1 - 10^{-10}$. }
\label{powerTab}
\end{table}

This scaling result is  not peculiar to five dimensional black holes, but in fact holds for any (odd) dimension.
 To see this we have determined numerically the scaling of $\Delta \mathcal{C}_\mathcal{V}$ near extremality \footnote{in the static limit, $S$ and $V^{(D-2)/(D-1)}$ behave in the same manner, so it is only the extremal limit that distiniguishes entropy from volume} in a number of higher odd dimensions, shown in Table~\ref{powerTab}. In all cases we see that the scaling matches precisely that derived from $V^{(D-2)/(D-1)}$. 
 
We thus find the intriguing result that the complexity of formation scales as  $V^{(D-2)/(D-1)}$, capturing two distinct scaling behaviours in the static and near-extremal limits as in \eqref{volScale}.  Interestingly this power is the same power that relates the entropy \eqref{entropy} to thermodynamic volume \eqref{vol} in the static solutions. However, while this means the scaling can be expressed in terms of either $S$ or $V^{(D-2)/(D-1)}$ near $r_-/r_+ \to 0$, it is only the volume that captures the correct scaling behaviour for all values of $r_-/r_+$ --- see Eqs.~\eqref{entScale} and~\eqref{volScale}.  

\textbf{Complexity equals action.}---We have now demonstrated that in the CV proposal it is the thermodynamic volume and not the entropy that characterizes the complexity of formation for large black holes. It is   natural to ask whether this behaviour is universal to both complexity proposals, or if it is a peculiar behaviour associated with the CV proposal. Here we show that the same feature emerges for the CA proposal.

In the CA proposal, the complexity of the CFT state at boundary time $t$ is given by the value of the gravitational action evaluated on the WDW patch of spacetime
\be 
\mathcal{C}_\mathcal{A}(\Upsilon) = \frac{I_{\rm WDW}}{\pi} \, .
\ee
The WDW patch is defined as the domain of dependence of the bulk Cauchy slice that intersects the boundary at the given timeslice $\Upsilon$. The geometry of this patch for the rotating black holes is shown in Fig.~\ref{penrose}. There are a number of non-trivial contributions to the action arising in this computation, including joint contributions at the future/past meeting points of the null sheets of the WDW patch, joint and boundary terms at the regularization of the patch near infinity, and a null boundary counterterm along the null sheets of the WDW patch. A full account of these terms will be presented elsewhere~\cite{UsFuture}, but it suffices to say that the computation is morally similar to the case of charged black holes~\cite{Carmi2017}.

The result of this analysis is that the complexity of formation in the CA proposal is given by
\be 
\Delta \mathcal{C}_\mathcal{A} = \frac{I_{\rm WDW} - 2 I_{\rm AdS}}{\pi}
\ee
with
\begin{widetext}
\begin{align}
\pi \Delta \mathcal{C}_\mathcal{A} =& \frac{\Lambda \Omega_{2N+1}}{2 (N+1) (2N+1) \pi G_N} \bigg[
\int_{r_{m_0}}^{\infty} dr r^{2N+1} \left( g(r)^2 h(r)  - \frac{r}{1 + r^2/\ell^2} \right)  - \int_0^{r_{m_0}} dr \frac{r^{2(N+1)}}{1 + r^2/\ell^2}  \bigg] -\frac{\Omega_{2N+1} (r_{m_0})^{2N+1}}{2 \pi G_N (2N+1)}
\nonumber\\
& -\frac{\Omega_{2N + 1}}{4 \pi G_N} (r_{m_0})^{2N} h(r_{m_0}) \log \ell_{\rm ct}^2 \Theta(r_{m_0})^2 |f(r_{m_0})^2|  
- \frac{\Omega_{2N+1}}{2 \pi G_N} \int_{r_{m_0}}^{\infty} dr  \, r^{2N} \left[  h(r) \frac{\Theta'}{\Theta} + 1 \right]
\end{align}
\end{widetext}
where
\begin{align}
\Theta  = \frac{1}{f(r) g(r)} \left[\frac{2N}{r} + \frac{h'}{h} \right]\, .
\end{align}
Here the constant $\ell_{\rm ct}$ comes from a counterterm on the null boundaries. Such a term is not required for a well-posed variational problem, but is required to ensure the final result does not depend on the parameterization of the null generators of the WDW patch~\cite{Lehner2016}, and moreover has been shown to be important for reproducing certain required properties of complexity in some situations~\cite{Chapman:2018dem, Chapman:2018lsv, Agon:2018zso, Alishahiha:2018lfv}.  The parameter $r_{m_0}$ is the value of $r$ at which the future/past tips of the WDW patch meet. It is determined by solving the equation $r^*(r_{m_0}) = 0$ where 
\be 
r^*(r) = \int_\infty^r \frac{g^2(\tilde{r}) h(\tilde{r})}{\tilde{r}} d \tilde{r} 
\ee
is the tortoise coordinate. 

\begin{figure}
\centering
\includegraphics[width=0.4\textwidth]{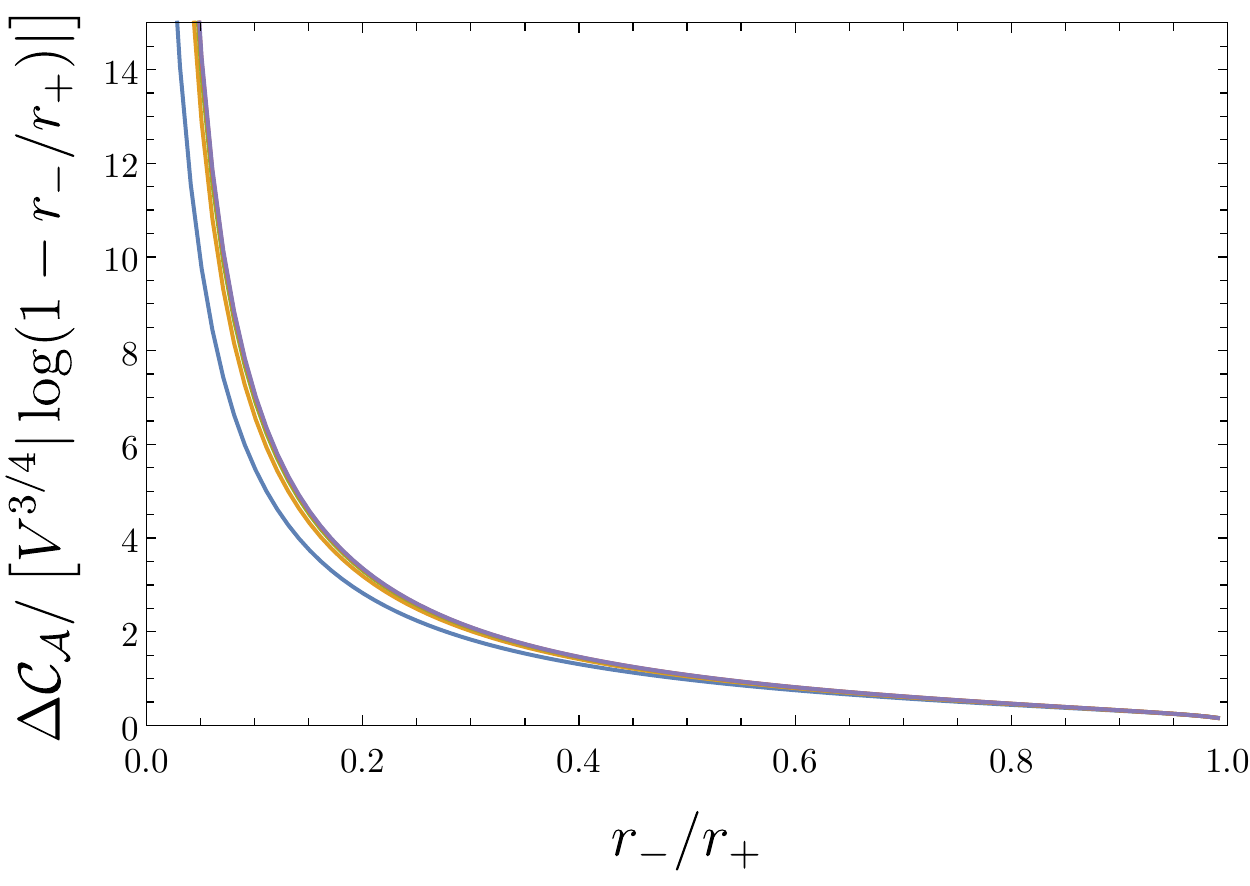}
\caption{A plot showing the CA complexity of formation normalized by the thermodynamic volume as a function of the ratio $r_-/r_+$ in five dimensions. The plot shows curves for fixed $r_+/\ell = 10, 10^2, 10^3, 10^4, 10^5, 10^6$ and $10^7$, however after $r_+/\ell = 1000$ the curves are visually indistinguishable. Here we have set $\ell_{\rm ct} = \ell$.}
\label{CAplot}
\end{figure}

The most difficult part of the CA computation is the determination of $r_{m_0}$. In some instances, particularly in the limit $r_-/r_+ \to 0$, accurate determination of this parameter requires hundreds of digits of precision in the numerics. This technicality has limited our ability to probe the behaviour of the complexity of formation within the CA conjecture as broadly as the CV conjecture. However, we show in Fig.~\ref{CAplot} the result of the action computation in five dimensions. The plot makes clear that the thermodynamic volume controls the scaling of $\Delta \mathcal{C}_\mathcal{A}$ for large black holes, just as in the CV conjecture.  While it was possible to compute the behaviour in various higher dimensions for the CV case, this is more difficult in the CA scenario. Nonetheless, we have confirmed the scaling with thermodynamic volume in seven dimensions, which suggests the same trend holds in general for CA.

\textbf{Discussion.}---We have shown here for the first time that the thermodynamic volume plays a natural role in both the CA and CV conjectures. Reinstating units, the complexity of formation of large black holes obeys the same scaling as the thermodynamic volume 
\be \label{DCbehave} 
\Delta \mathcal{C} = {\Sigma}_{\rm g} C_T \left(\frac{V}{V_{\rm AdS}}\right)^{\frac{D-2}{D-1}} 
\ee
where $V_{\rm AdS} = \ell^{D-1}$, $\Sigma_{\rm g}$ is a factor that depends on the specific metric, dimension, etc. but not on the size of the black hole, and 
$C_T \sim \ell^{D-2}/G_N$ is the central charge of the CFT.

This proposal reproduces known results for static black holes, as in those cases the thermodynamic volume is not independent from the entropy, $S \sim V^{(D-2)/(D-1)}$, and the above can be recast in terms of the entropy in those cases.  However, for rotating black holes the volume and entropy are independent and it becomes clear that it is~\eqref{DCbehave} that captures the correct behaviour, and not an analogous expression involving the entropy. Our result also reproduces the behaviour of the complexity of formation for gravitational solitons~\cite{Andrews:2019hvq}, which are horizonless geometries that possess thermodynamic volume but no entropy. 

The thermodynamic volume has been conjectured~\cite{Cvetic:2010jb} to obey a `reverse' isoperimetric inequality:
\be 
\mathcal{R} \equiv \left(\frac{(D-1) V}{\Omega_{D-2}} \right)^{1/(D-1)} \left(\frac{ \Omega_{D-2}}{ 4 G_N S} \right)^{1/(D-2)} \ge 1 \, .
\ee
The inequality is saturated by (charged) Schwarzschild-AdS spacetimes. Assuming the relationship~\eqref{DCbehave} is general, the reverse isoperimetric inequality becomes the statement
\be\label{isoBound} 
\Delta \mathcal{C} \ge \beta_D S
\ee
where $\beta_D$ is a positive constant that can be easily worked out from the above. This means that the complexity of formation for large black holes is \textit{bounded from below} by the entropy (equivalently, the number of degrees of freedom).
 
What we have said so far concerns the complexity of formation. Before closing, let us remark that there is also a connection between the late time growth of complexity and thermodynamic volume, again for large black holes. For the rotating black holes considered here this relationship works out to be~\cite{UsFuture}
\be 
\dot{\mathcal{C}} = N_{\mathcal{A}, \mathcal{V}} P \Delta V
\ee
where $\Delta V$ is the difference between the thermodynamic volume of the inner and outer horizons and $N_{\mathcal{A}, \mathcal{V}}$ is a proportionality constant whose numeric value depends on whether one uses the CV or CA conjecture. The implication of this is that not only does the thermodynamic volume control the complexity of formation, but we see here that it also controls the late-time growth.


Our results for the complexity of formation draw a clear and simple connection between thermodynamic volume and holographic complexity. A better understanding of complexity in the holographic dictionary would then lead to a simple and direct holographic interpretation of thermodynamic volume and vice versa. Going forward, it will be important to assess the validity of our proposal~\eqref{DCbehave} as broadly as possible. Exploring the properties of complexity of formation in other spacetimes where $S$ and $V$ are independent would contribute additional evidence toward the generality of the relationship, or could constitute a counter-example from which its possible limitations could be assessed.

\textbf{Acknowledgements.}---We thank Hugo Marrochio for useful discussions. This work was supported in part by the Natural Sciences and Engineering Research Council of Canada. The work of RAH is supported by the Natural Sciences and Engineering Research Council of Canada through the Banting Postdoctoral Fellowship program.  HKK acknowledges the support of NSERC Discovery Grant RGPIN-2018-04887.

\bibliographystyle{apsrev4-1} 
\vspace{1cm}
\bibliography{myrefs}

\onecolumngrid  \vspace{0.6cm} 
\begin{center}  
{\Large\bf Supplemental Material} 
\end{center} 
\appendix  

\section{Determination of $\Delta \mathcal{C}_\mathcal{V}$ scaling}

In this supplement, we describe for the five dimensional case how the scaling of $\Delta \mathcal{C}_\mathcal{V}$ was determined. The same method was used in higher dimensions, and also for the action complexity with the only difference being the more complicated expressions.

To evaluate the complexity of formation in the CV proposal, we must understand the behaviour of the following integral:
\be 
I(\alpha, \epsilon) = \int_0^1 \left[\frac{1}{u^4 \alpha^4} \sqrt{\frac{\alpha^2 + u^2(\epsilon-1)^2(\alpha^2 + 2 - 2 \epsilon + \epsilon^2)}{(u^2-1)(u^2(\epsilon-1)^2 - 1)(1 + u^2(\alpha^2 + 2 - 2 \epsilon + \epsilon^2)) }} - \frac{1}{u^4 \alpha^3 \sqrt{1 + \alpha^2 u^2}}\right] du
\ee
where we have defined the quantities
\begin{align}
r = \frac{r_+}{u} \, , \quad \alpha = \frac{\ell}{r_+} \, , \quad \epsilon = 1- \frac{r_-}{r_+} \, .
\end{align}
In terms of the above, the complexity of formation is written
\be 
\Delta \mathcal{C}_\mathcal{V} = \frac{2 \Omega_{D-2} \ell^4}{G_N R} \left[I(\alpha, \epsilon) - \frac{1}{3 \alpha^2} \left(2 \alpha (\alpha - \sqrt{1 + \alpha^2}) + \sqrt{1 + \frac{1}{\alpha^2}} \right) \right] \, .
\ee
The objective is to understand how $I(\alpha, \epsilon)$ behaves as a function of $\alpha$ for small $\alpha$ when $\epsilon$ is close to zero. That is, we are interested in large black holes near to extremality. (As discussed in the main text, the only subtlety comes near extremality, as the solutions reduce to the Schwarzschild-AdS geometries when $r_-/r_+ \to 0$ for which the scaling is known).

Unfortunately, we have been unable to obtain a (useful) exact result for the integral, nor have we succeeded in obtaining an asymptotic expansion for small $\alpha$. Instead, we have studied this problem numerically. Our method was as follows. Supposing that $I(\alpha, \epsilon) \sim \alpha^{-\gamma}$ for some power $\gamma$, we consider the combination
\be\label{combo} 
R(\beta) \equiv \alpha^\beta I(\alpha, \epsilon)  \sim \alpha^{\beta-\gamma} \, .
\ee
We then study the logarithm of this object, treated as a function of $\beta$. For each choice of $\beta$, we evaluate $\log R(\beta)$ for several (small) values of $\alpha$. We fit these results with a linear model and extract the slope. Exploring the $\beta$-parameter space, we search for the value of $\beta$ for which the slope determined in this way vanishes. This value corresponds to $\beta = \gamma$. Once such a determination has been made, we can then perform additional convergence tests on the combination~\eqref{combo} to ensure that the behaviour is correct. We have carried out this procedure using Mathematica. Our numerical integrals have been computed using a working precision of 500 with a precision goal of 50. The values in the table in the main text were obtained by doing this for $\alpha$ between $10^{-20}$ and $10^{-10}$, with $\epsilon = 10^{-10}$.  The results in 5, 7, and 9 dimensions were independently cross-checked using Maple.

\end{document}